\documentclass[natbib209]{emulateapj}

\usepackage{graphicx}

\shortauthors{Bournaud et al.}

\begin{document}

\title{The thick disks of spiral galaxies as relics from gas-rich,
turbulent, clumpy disks at high redshift}

\author{Fr\'ed\'eric Bournaud}
\affil{CEA, IRFU, SAp, 91191 Gif-sur-Yvette, France.}

\author{Bruce G. Elmegreen}
\affil{IBM Research Division, T.J. Watson Research Center, P.O. Box
218, Yorktown Heights, NY 10598, USA.}

\author{Marie Martig}
\affil{CEA, IRFU, SAp, 91191 Gif-sur-Yvette, France.}

\begin{abstract}
The formation of thick stellar disks in spiral galaxies is studied. Simulations of gas-rich young galaxies show formation of internal clumps by gravitational instabilities, clump coalescence into a bulge, and disk thickening by strong stellar scattering. The bulge and thick disks of modern galaxies may form this way. Simulations of minor mergers make thick disks too, but there is an important difference. Thick disks made by internal processes have a constant scale height with galactocentric radius, but thick disks made by mergers flare. The difference arises because in the first case, perpendicular forcing and disk-gravity resistance are both proportional to the disk column density, so the resulting scale height is independent of this density. In the case of mergers, perpendicular forcing is independent of the column density and the low density regions get thicker; the resulting flaring is inconsistent with observations. Late-stage gas accretion and thin disk growth are shown to preserve the constant scale heights of thick disks formed by internal evolution. These results reinforce the idea that disk galaxies accrete most of their mass smoothly and acquire their structure by internal processes, in particular through turbulent and clumpy phases at high redshift.
\end{abstract}

\keywords{ISM: structure --- Galaxy: disk --- galaxies: formation
---galaxies: high-redshift}

\section{Introduction}

In the cosmological model of galaxy formation, gas-rich
mergers are necessary to preserve disks and build spiral galaxies
\citep[e.g.,][]{robertson06}. These mergers also scatter disk stars
and produce the thick disk component \citep{quinn93,walker96}. There is however increasing evidence that most of the baryonic mass enters in cold flows, which are smoother and less disruptive than mergers \citep{agertz09, ocvirk08, keres09}. Cold flow 
accretion is expected to be most active at $z\sim 2$ \citep{dekel09}. Observations of galaxies around that epoch show a large population of rotating disks that are clumpy and turbulent \citep{ee05, genzel08, shapiro, elmegreen09a,elmegreen09b}. The clumps in these galaxies have masses up to a few $10^8$~M$_{\sun}$, sizes of around a kpc, and likely formed internally by gravitational instabilities \citep{elmegreen07}. Cold flows have an advantage over mergers as an explanation for high-redshift clumpy galaxies (Agertz et al. 2009: Ceverino, Dekel \& Bournaud 2009). This is because the formation of the observed clumps requires a relatively high disk velocity dispersion for the ambient Jeans mass to be large. Then a high fraction of the baryons has to be in the primordial disk to keep the system unstable: only relatively smooth gas accretion can do that \citep{BE09,DSC09}.

The origin of thick disks then comes into question. If most of the mass assembly is from smooth gas flows and rapid internal evolution, then thick disks have to form by internal processes, such as stellar scattering off of clumps and disk heating by gravitational instabilities that make clumps \citep[][hereafter BEE07]{BEE07}. 
Clumpy disk evolution can form classical bulges \citep{noguchi99, immeli04, EBE08}, and exponential disks (Bournaud, Elmegreen \& Elmegreen 2007, hereafter BEE07), but there are no distinct features in these remnants that exclusively point to clump-driven evolution. Perhaps thick disk formation by clump scattering is a more revealing remnant in modern galaxies.

Here we explore the properties of thick disks formed internally in unstable, gas-rich, clumpy disks. We compare with simulations of merger-induced disk thickening. We find that thick disks formed internally have a nearly constant scale height with radius, as observed for real galaxies. This is unlike the case for merger-induced thick disks, which always flare. We also investigate how much the thick disk changes during later gas accretion when the thin disk forms, and find that the scale height decreases, but remains constant with radius.

\section{Properties of observed thick disks}

\begin{figure}[!ht]
\centering
\includegraphics[width=3.2in]{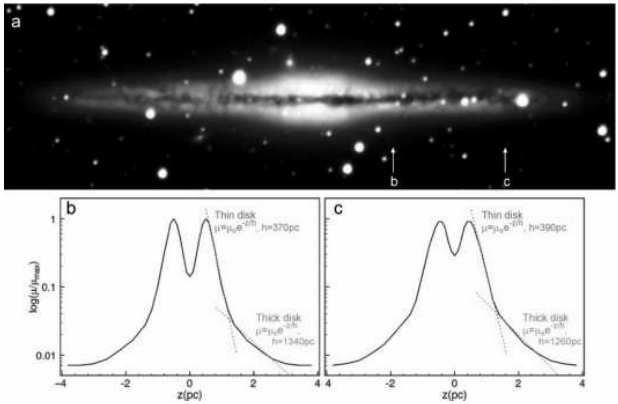}
\caption{Edge-on view of the Milky Way analogue NGC~891 in the $R$-Band from 2Mass data
(a). Vertical luminosity cuts in the $3<r<4$~kpc (b) and $6<r<8$~kpc
(c) radial bins, avoiding any foreground star, show the thin mid-plane dust lane, the thin stellar
disk, and the thick stellar disk. The thick disk has a relatively
constant scale-height. Even the deepest HST/ACS data for NGC~891 can be
fitted with a thin+thick disk model having uniform scale-height
\citep{ibata09}.}\label{fig:1}
\end{figure}

The primary remnants of the earliest stage of galaxy formation are the
bulge and thick disk components.  The thick disk in the Milky Way
contains no stars younger than 8~Gyr \citep{gilmore85,
reddy06}.  Thick disks are ubiquitous \citep{dalcanton02,seth05}, and contain 10\% to 25\% of the baryonic
mass of spiral galaxies. In the disk alone, thick-to-thin mass ratios
range from 10\% to 50\% (Yoachim \& Dalcanton 2006, hereafter YD06); in
the Milky Way it is $20-25$\% \citep{gilmore-reid83,chen01,robin03}.

The structural properties of thick disks that are relevant for
comparisons with our models are:
\begin{itemize}
\item The thick-disk ${\rm sech}^2$ scale height, $z_0$, is generally between 1.4 and 2.8~kpc, with a large scatter (Dalcanton \& Bernstein 2002, YD06).  Equivalent exponential-fit scale
heights, $h_z$, are about half these values.

\item $z_0$ is about constant
with radius (YD06). In the Milky Way \citep{robin03} and other galaxies
\citep[e.g.,][]{ibata09}, $h_z$ also does not have significant
variations until $R_{25}$ and beyond. This constancy of $h_z$ is
illustrated here in Figure~\ref{fig:1}, where vertical profiles from
the inner and outer parts of NGC~891 are presented. A constant $h_z$ implies that thick disks viewed edge-on have disky isophotes. Signs of boxiness were sought but not found (Dalcanton \& Bernstein 2002).

\item Thick and thin disks have relatively similar radial scalelengths for
most galaxies (YD06).
\end{itemize}

Bulges and thick disks have enhanced $\alpha$ elements compared to Fe \citep[e.g.,][]{lecureur07,zoccali07}, suggesting a star 
formation timescale shorter than 1~Gyr. Other chemical similarities between the bulge and the thick disk of the Milky Way were discussed by \citet{melendez08}.

\section{Simulations}
\subsection{Unstable disks at high-redshift and thin disk growth}
The simulations used here are similar to those in BEE07, but with an
increased resolution to see the vertical disk structure. The $N$-body particle-mesh code uses a sticky particle scheme to model cold
gas dynamics. The spatial resolution (softening size) is 35~pc. Two million particles
are used for each component of gas, stars, and dark matter. We ran 3
models:
\begin{itemize}
\item Model 1 starts with a stellar mass of
2$\times$10$^{10}$~M$_{\sun}$ and a gas fraction of 60\%.
\item Model 2 starts with a stellar mass of
3$\times$10$^{10}$~M$_{\sun}$ and a gas fraction of 50\%.
\item Model 3 starts with a stellar mass of
4$\times$10$^{10}$~M$_{\sun}$ and a gas fraction of 40\%.
\end{itemize}
The models start with idealized, uniform disks having masses that are
plausible for $z$$\sim$2 progenitors of Milky Way-like spirals. They rapidly acquire a phase space distribution (morphology and kinematics) that is similar to that in cosmological simulations of clumpy disk formation (Agertz et al. 2009; Ceverino et al. 2009) and also consistent with that of observed clumpy galaxies (see BEE07 and Bournaud et al. 2008). Thus the present models should be adequate for studies of disk evolution during and after the clumpy phase. 

The initial disks have constant surface density for both gas and stars.
The external radius is 6~kpc and the initial scale height is
$h_z=500$~pc. The ratio of dark matter to baryon mass is 1:2 inside the initial disk radius. The models are first run for 1~Gyr in full
isolation in order to identify the effects of internal evolution. 

After this time, additional gas is allowed to accrete for another 6~Gyr. This is supposed to model the later growth of the thin disk. The gas
accretion rate is 10~M$_{\sun}$~yr$^{-1}$, which increases the initial
galaxy mass by a factor 2 to 3. Accreted particles are positioned at a
large radius (15~kpc), close to the disk plane:  they are
initially given a vertical distribution with an exponential profile and a scale
height of 1.5~kpc, and later-on settle in a thinner disk. Accreted particles are not all present in the simulation initially, they are continuously added at the $r$=15~kpc boundary to model continuous accretion down to low redshifts, as in \citet{BC02}. Each is given a rotational velocity such that
it has the same angular momentum as an average particle in the thick
disk at 1~Gyr. This similarity of angular momenta gives the thick and
thin disks similar radial scale lengths. Thus the additional gas is accreted from a thick
layer in these models, but is relatively stable and naturally settles
to a thin disk.

\subsection{Minor mergers}
Many simulations of minor mergers have been published
\citep{walker96,VH08,read09}, in addition to cosmological simulations with with numerous satellites accreting and merging \citep{abadi}. We performed new merger models to compare with the internal evolution using the same code and assumptions.
We ran four idealized simulations of single minor mergers and a
cosmological simulation where several minor mergers occur with mass
ratios and orbits prescribed by a $\Lambda$-CDM model.

The four individual minor mergers were chosen to have a mass ratio of
10:1. The primary galaxy is modeled with one million particles for each
component (gas, stars, dark matter). Its stellar mass is
4$\times$10$^{10}$~M$_{\sun}$ and its gas mass is
1$\times$10$^{10}$~M$_{\sun}$. It starts with an exponential disk of
radial scale-length 4~kpc, truncated at 15~kpc, and an exponential
scale-height of 500~pc. 20\% of the stellar mass is in a central bulge
with a Hernquist profile of scale-length 500~pc. The dark matter halo
has a Hernquist profile with a scale-length of 7~kpc. The dark
matter-to-baryon mass ratio is 1:2 inside the disk radius. The
companion galaxy has all masses divided by a factor of 10, and all
sizes divided by 3.5. The orbital inclination with respect to the
target disk plane is 35 degrees. The impact parameter is 30~kpc. The
velocity at an infinite distance is either 150 or 220~km ~s$^{-1}$,
each being simulated for a prograde and a retrograde orbit, thus giving
four models. The resolution and softening are the same as in the unstable disk models

Our second type of merger model includes several satellite galaxies
taken from a large-scale cosmological simulation (see Martig et al. 2009a,b). The primary galaxy in this case uses
4.7$\times$10$^6$ stellar particles, 3.2$\times$10$^6$ gas particles,
and 5.7$\times$10$^6$ dark matter particles. The softening length for
force calculations is 150~pc in this model, larger than in the other simulations, but the final thick disk scaleheight is still well resolved. The disk is initially thin at redshift
$z=1.2$, and it contains 30\% of its mass in the form of gas. Over
time, the galaxy interacts with several dwarf satellites, the most
massive of which have mass ratios of 11:1, 13:1, 15:1, and 18:1. The
final stellar mass in the primary is 8$\times$10$^{10}$~M$_{\sun}$. We
analyze its thick disk at $z=0$.

\section{Results}
\subsection{Thick disks from instabilities and giant clumps
at high redshift}

\begin{figure}
\centering
\includegraphics[width=1.9in]{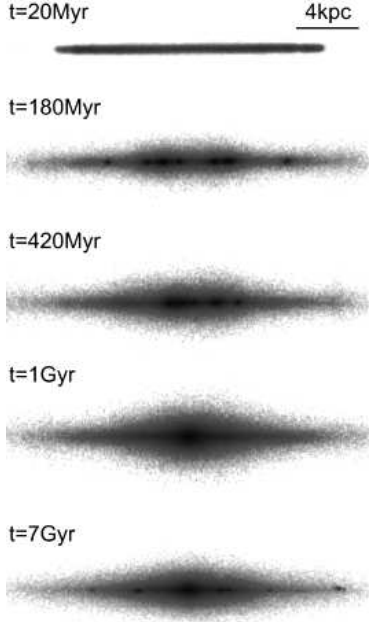}
\caption{Model~2 of a gas-rich gravitationally-unstable disk generating
giant clumps of gas and stars, similar to clumpy galaxies at $z>1$ (see
also BEE07). These edge-on views of the gas+star mass density show
rapid thickening of the disk by the clumps. The clumps finally coalesce
into a central bulge within a Gyr. Subsequent gas infall forms a stable
thin disk inside the thick disk. The thick disk scale-height decreases
a little when this happens, but the separate thick disk component remains.
}\label{fig:2}
\end{figure}

\begin{figure}
\centering
\includegraphics[width=3in]{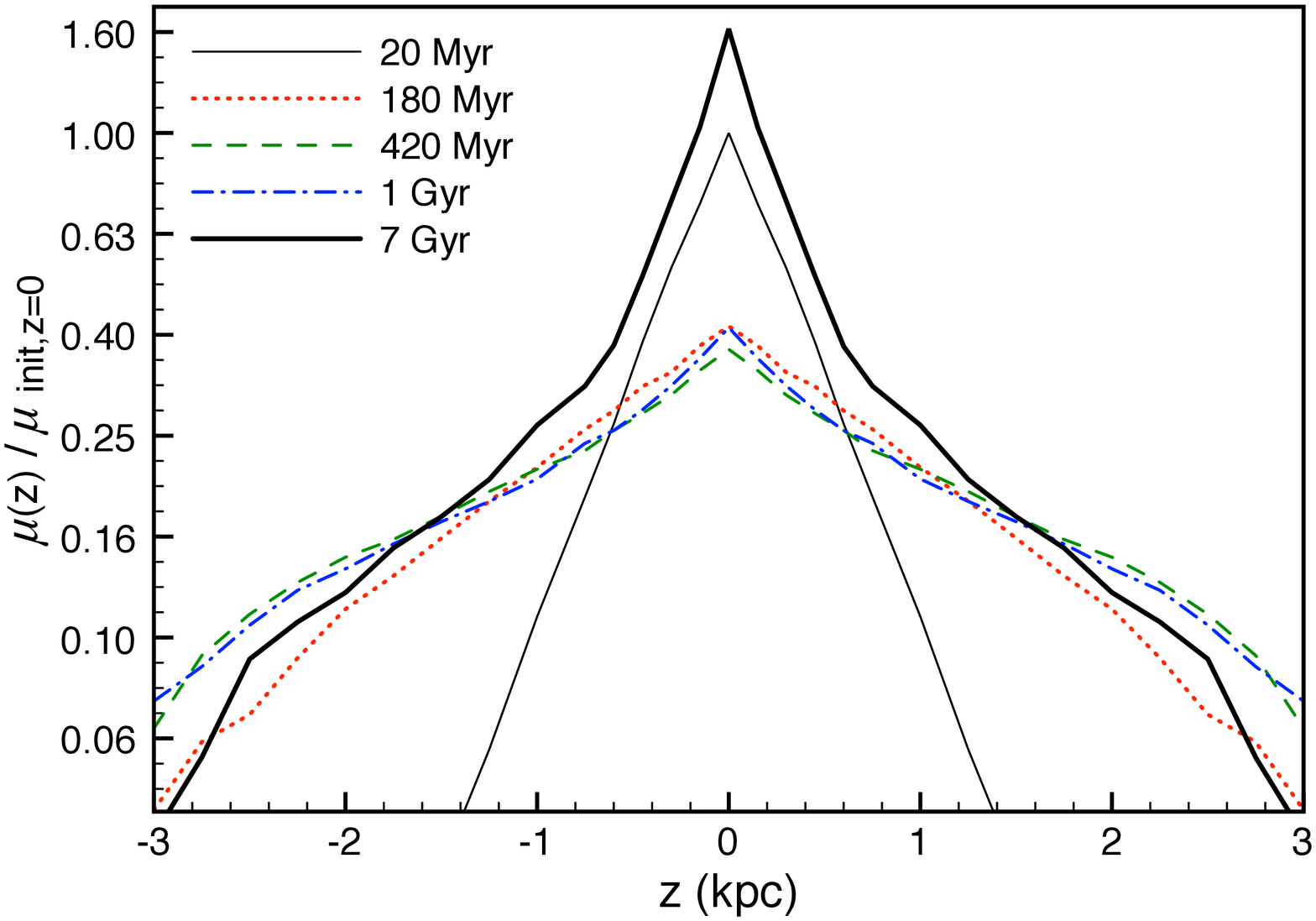}
\caption{Vertical line-of-sight stellar density profiles for the
systems shown on Figure~2, at $r$=5~kpc (see Fig.~4 for radial
variations), normalized to the mid-plane density of the first snapshots
and plotted in log scale. The initial thin but unstable disk is rapidly
stirred and thickened by gravitational instabilities and clumps. A
thick disk is formed within a Gyr, and a faint thin disk forms inside
of this from left-over gas. Continued external accretion causes a more
prominent young thin disk to form, and the thick disk shrinks a little
when this happens.}\label{fig:3}
\end{figure}

The initial Gyr in our models produces giant star-forming clumps with
masses up to a few $10^8$~M$_{\sun}$. The clumps eventually merge into a bulge (see also BEE07). A thick disk
forms quickly because of the same gravitational instabilities that form
the clumps. That is, the instabilities scatter stars and gas to high
velocity dispersions, and the clumps that are formed by the
instabilities continue to scatter stars and increase the dispersion for
as long as they are present.

The exponential scale-heights in the stellar disks were measured after
the clumpy phase, when the clumps have either dissolved or coalesced in
a bulge. The scale-height is $\sim2$ to 2.5~kpc. It is constant with
radius (Fig.~\ref{fig:2}), having only a minor flare in the outer
regions. That is, at a radius of 4 times the disk radial scale-length,
the scale-height increases to $\sim$3~kpc.

The slow addition of gas after 1 Gyr causes the thick disk to shrink
toward the mid-plane, as expected \citep{EE06}. In our models, the
scale-height of the thick disk decreases from $\sim2$ or 2.5 kpc  to
$\sim1.5$~kpc (Fig.~3 and 4). In all models, the thick disk remains a separate component and does not become a low-density tail to the thin disk (Fig.~3). The thick disk also preserves its constant scale-height with radius during the accretion (Fig.~4).

\begin{figure}
\centering
\includegraphics[width=3in]{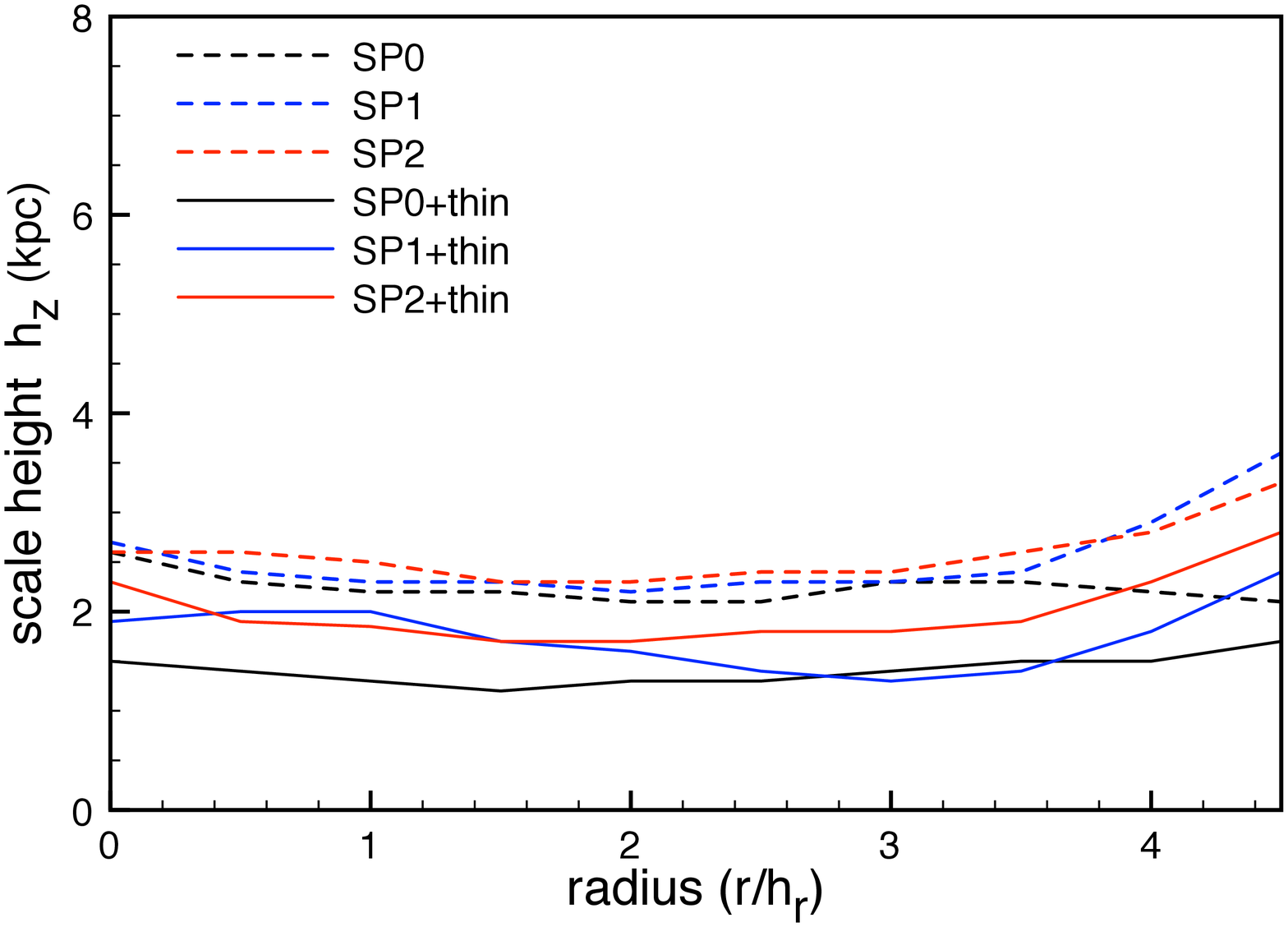}\\
\includegraphics[width=3in]{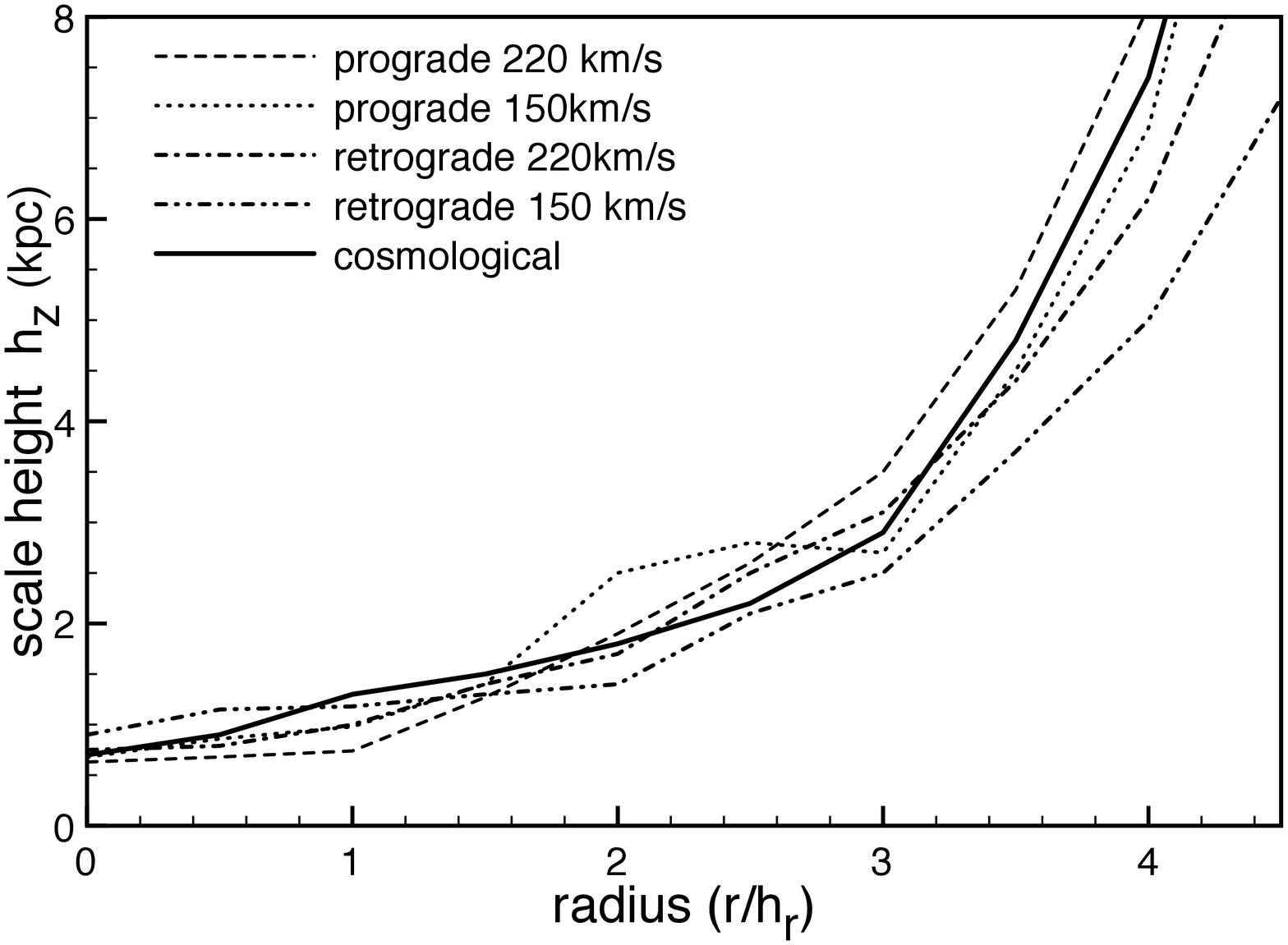}
\caption{Radial variations of the thick disk exponential scale-height
$h_z$ are shown as a function of the radius $r$ in units of the radial
exponential scale-length $h_r$.  On the top, models of thick disk
formation by internal evolution are shown after 1 Gyr (dashed lines)
and after subsequent thin disk accretion (dotted lines). On the bottom,
models of thick disk formation by 1:10 mergers are shown in four cases (broken lines) and in a cosmological case (solid line). The scale height is about constant for models with internal disk evolution, but it rapidly increases with radius for models with mergers.}\label{fig:4}
\end{figure}

\subsection{Thick disks from minor mergers}
The radial profiles of the thick-disk scale-heights in our minor merger
models are also shown in Fig.~\ref{fig:4}. They have no large
thickening in the inner regions ($r$$\simeq$0) and the outer disk
becomes extremely thick, with scale-heights of 5~kpc and more. Minor
mergers produce {\it flared} disks instead of uniformly thick disks.
This property is not exclusive to our code or initial conditions: other
published simulations of disk thickening by minor mergers show the same
thing \citep{walker96,VH08,read09}. The strong flaring often results in
boxy isophotes for edge-on thick disks \citep{VH08}.

Flaring is inevitable in merger models because the dense inner regions
have stronger disk forcing compared to companion forcing and the
perpendicular response is small. In contrast, low-density outer regions
have lower local disk forcing and stronger companion forcing, so the
disk responds more. Even if an interaction provides the same amount of
kinematic heating to the whole disk, the resultant scale height will
vary as the inverse of the surface density. Minor mergers would also
add more energy and mass to the outer disk than the inner disk because
of the disruption of the small galaxy. Self-driven disk thickening by
internal clumps does not produce much flaring because the perpendicular
forcing and the disk gravity that resists this forcing are both
proportional to the surface density.

\section{Comparison to observed thick disks}
Observed thick disks have a relatively constant scale height from the
central regions to the outskirts of a galaxy. Minor merger models do
not have this property, and they also lack any thick component in the
central regions. Models where thick disks form by gravitational
instabilities and clump scattering do have this property, however.
Therefore, the disky shapes of thick disks are explained exclusively by
the instability model.

The scale-heights of our thick disk models are consistent with the
observed values of 1 to 2~kpc for typical bright spirals, even after
the thick disk shrinks from the growth of an underlying thin disk. For
Milky Way galaxy masses, our models end up with about one-third of the
initial baryons in a bulge, one-third in the thick stellar disk, and
one third in gas that will contribute to the subsequent thin disk.
After further thin disk accretion, the final masses for the bulge,
thick disk, and thin disk become $2-3\times 10^{10}\;M_\odot$,
$2-3\times 10^{10}\;M_\odot$, and $8-9\times 10^{10}\;M_\odot$,
respectively. Thus the thick-to-thin disk mass ratio is 30\%, roughly
typical of modern spiral galaxies.

The thick disks formed in our models are relatively massive, $2-3 \times 10^{10}\;M_\odot$, but consistent with many observed spiral galaxies. A larger set of lower-resolution simulations in BEE07 showed that the fraction of the clump mass that ends up in the disk can vary largely, so our model can account for a wide range of thick disk masses. In this case an anti-correlation between the bulge mass fraction and the thick disk mass fraction is predicted. That is, clump stars that do not scatter into the thick disk will end up in the bulge. YD06 do observe that late-type spirals (which have low bulge fractions) have higher thick-to-thin disk mass ratios than early-type spirals (which have large bulge fractions).

Our models suggest that minor mergers are not the main formation
mechanism for thick disks. Still, they occur for real galaxies over a
wide range of redshifts, and they probably also occurred when the old
thick disks formed. Their effect on the thick disk seems to have
been minor though. One problem minor mergers have is that the gas
fraction of disk galaxies is high at high redshift ($z>1$;
Daddi et al. 2008), and a high gas fraction can prevent minor mergers from thickening the stellar disk \citep{moster09}. A problem at low redshift $z<1$ is that the thick disk is already in place, so minor mergers cannot impact its structure much \citep{VH08}.  Also, the mass of a young disk can be larger than the mass of a minor companion, and the disk is closer to its own stars than the companion is, so the instantaneous gravitational force on the disk from its own stars is generally much larger than the force from a companion. 
Thus internal processes should dominate thick disk formation if young disks are clumpy. This is the essential point of the present models: the gravitational potential in young disks is extremely clumpy and time-variable because of violent instabilities and the presence of spiraling massive clumps.

The stellar disks of spiral galaxies do have some outer flaring, but it
begins only at large radii and it affects mainly the thin disk
\citep[e.g.,][for the Galaxy]{alard00,momany}. This situation is
consistent with a picture in which the thick disk forms mostly by
internal processes at high redshift, and the primary effect of
interactions is to warp and flare the outer thin disk at lower
redshift. Stellar streams within the thick disk could also result from
dwarf companions \citep{gilmore02}.

An additional consideration is that the young clumpy phase of a galaxy
should form only one thick disk component that has a single large
vertical velocity dispersion. If minor mergers form the thick disk,
then each merger could form a distinction component. The absence of any
significant vertical gradient of the stellar velocity dispersion (Moni
Bidin et al. 2009) suggests there is only one thick disk component.

Thick disk formation by internal forcing is also consistent with the
observed elemental abundances. Our models indicate that thick disks and
bulges form together within a time span of $\sim$1~Gyr. This explains
their elevated $\alpha/Fe$ abundances and other abundance similarities
\citep{melendez08}. If minor mergers were the main drivers of thick
disk formation, then thick disks would continue to grow below $z<1$ (as illustrated by our cosmological model). In that case, the age
range, metallicity range, and formation timescale for the thick disk
would differ from the observations. Other models of thick disk
formation by internal evolution, like cluster disruption \citep{kroupa02} or radial mixing \citep{roskar,schonrich-binney09}, would also
imply an unacceptable continuity of stellar ages in the thick disk.

\section{Conclusion}
There is increasing evidence that many disk galaxies at high redshift
undergo a clumpy and turbulent phase that can drive rapid
internal evolution. To understand if this is the main mode of spiral
galaxy formation, remnants of this phase should be idenfitied in the
oldest components of today's galaxies. The two most obvious of these
old components are the bulge and thick stellar disk. Previous
models showed how the clumps formed by gravitational instabilities could coalesce into a bulge, but bulges can form by galaxy mergers as well. Here we showed how the clumps, and the instabilities that formed them, can also make the thick disk, while
minor mergers fail to explain some observed properties of thick disks.

The standard model in which minor mergers stir an existing thin disk
and deposit stars far from the plane has a problem: it inevitably makes
flared thick disks because the vertical forcing is independent of the
local disk column density. Such flaring is contrary to observations.
Our model produces a uniform disk thickness because the forcing is
proportional to the column density. This property is preserved during
subsequent thin disk formation by continued slow accretion. In
addition, our model reproduces the mass fractions of thick and thin
disks and explains, in general terms, the chemical similarities of the
Milky Way bulge and thick disk. 

The observed properties of thick disks suggest that today's spiral galaxies went through a clumpy phase with rapid internal evolution, and that clumpy disk galaxies at high redshift are progenitors of spiral galaxies.

\acknowledgements
Helpful comments from the referee are acknowledged. This work was performed using HPC resources from GENCI-CCRT (Grant 2009-042192), and supported by Agence Nationale de la Recherche under contract ANR-08-BLAN-0274-01.


\begin{thebibliography}{plain}

\bibitem[Abadi et al.(2003)]{abadi} Abadi, M.~G., Navarro, J.~F., Steinmetz, M., \& Eke, V.~R.\ 2003, \apj, 597, 21 

\bibitem[Alard(2000)]{alard00} Alard, C.\ 2000,
arXiv:astro-ph/0007013

\bibitem[Agertz et al.(2009)]{agertz09} Agertz, O., Teyssier, R., Moore, B. 2009,
MNRAS, 397, L64

\bibitem[Bournaud \& Combes(2002)]{BC02} Bournaud, F., \& Combes, F.\ 2002, \aap, 392, 83


\bibitem[Bournaud et al.(2007)]{BEE07} Bournaud, F.,
Elmegreen, B.~G., \& Elmegreen, D.~M.\ 2007, \apj, 670, 237 (BEE07)

\bibitem[Bournaud et al.(2008)]{bournaud08} Bournaud, F., et al.\ 2008, \aap, 486, 741

\bibitem[Bournaud
\& Elmegreen(2009)]{BE09} Bournaud, F., \& Elmegreen, B.~G.\ 2009, \apjl, 694, L158

\bibitem[Chen et al.(2001)]{chen01} Chen, B., et al.\ 2001, \apj, 553, 184

\bibitem[Ceverino et al.(2009)]{ceverino} Ceverino, D., Dekel, A., \& Bournaud, F. 2009,
arXiv0907.3271

\bibitem[Daddi et al.(2008)]{daddi08} Daddi, E., Dannerbauer,  H.,
Elbaz, D., Dickinson, M., Morrison, G., Stern, D., \& Ravindranath, S.\
2008, \apjl, 673, L21

\bibitem[Dalcanton
\& Bernstein(2002)]{dalcanton02} Dalcanton, J.~J., \& Bernstein, R.~A.\ 2002, \aj, 124, 1328

\bibitem[Dekel et al.(2009)]{dekel09} Dekel, A., et al.\ 2009, Nature, 457, 451

\bibitem[Dekel et al.(2009)]{DSC09} Dekel, A., Sari, R.,
\& Ceverino, D.\ 2009, \apj, 703, 785

\bibitem[Elmegreen \& Elmegreen(2005)]{ee05} Elmegreen, B., \& Elmegreen, D. 2005, ApJ, 627, 632

\bibitem[Elmegreen
\& Elmegreen(2006a)]{EE06} Elmegreen, B.~G., \& Elmegreen,
D.~M.\ 2006a, \apj, 650, 644

\bibitem[Elmegreen et al.(2007)]{elmegreen07} Elmegreen, D.~M.,
Elmegreen, B.~G., Ravindranath, S., \& Coe, D.~A.\ 2007, \apj, 658, 763

\bibitem[Elmegreen et al.(2008)]{EBE08} Elmegreen, B.G., Bournaud, F., \& Elmegreen, D.M. 2008, ApJ, 688, 67

\bibitem[Elmegreen et al.(2009a)]{elmegreen09a} Elmegreen, D.M.,
Elmegreen, B.G., Fernandez, M.X., \& Lemonias, J.J., 2009a, ApJ, 692,
12

\bibitem[Elmegreen et al.(2009b)]{elmegreen09b} Elmegreen, D.M., Elmegreen, B.G., Marcus, M.T.,
Shahinyan, K., Yau, A., \& Petersen, M. 2009b, ApJ, 701, 306

\bibitem[Forster Schreiber et al.(2009)]{FS09} Forster
Schreiber, N.~M., et al.\ 2009, ApJ submitted (arXiv:0903.1872)

\bibitem[Genzel et al.(2008)]{genzel08} Genzel, R., et al.\
2008, \apj, 687, 59

\bibitem[Gilmore
\& Reid(1983)]{gilmore-reid83} Gilmore, G., \& Reid, N.\ 1983, \mnras, 202, 1025

\bibitem[Gilmore et al.(1985)]{gilmore85} Gilmore, G., Reid, N., \& Hewett, P.\ 1985, \mnras, 213, 257

\bibitem[Gilmore et al.(2002)]{gilmore02} Gilmore, G., Wyse,
R.~F.~G., \& Norris, J.~E.\ 2002, \apjl, 574, L39

\bibitem[Ibata et al.(2009)]{ibata09} Ibata, R., Mouhcine, M.,
\& Rejkuba, M.\ 2009, \mnras, 395, 126

\bibitem[Immeli et al.(2004)]{immeli04} Immeli, A., Samland, M., Gerhard, O., \& Westera, P.
2004, A\&A, 413, 547

\bibitem[Keres et al.(2009)]{keres09}  Keresš, D., Katz, N., Fardal, M., Dav\'e, R., \& Weinberg,
D.~H.\ 2009, MNRAS, 395, 160

\bibitem[Kroupa(2002)]{kroupa02} Kroupa, P.\ 2002, \mnras, 330, 707

\bibitem[Lecureur et
al.(2007)]{lecureur07} Lecureur, A., Hill, V., Zoccali, M., Barbuy, B., G{\'o}mez, A., Minniti, D., Ortolani, S., \& Renzini, A.\ 2007, \aap, 465, 799

\bibitem[Martig et al.(2009a)]{martig09a} Martig, M., Bournaud,
F., \& Teyssier, R.\ 2009, IAU Symposium, 254, 429

\bibitem[Martig et al.(2009b)]{martig09b} Martig, M., Bournaud, F., Teyssier, R., \& Dekel, A.\ 2009, ApJ submitted (arXiv:0905.4669)

\bibitem[Mel{\'e}ndez et al.(2008)]{melendez08} Mel{\'e}ndez, J., et al.\ 2008, \aap, 484, L21

\bibitem[Momany et
al.(2006)]{momany} Momany, Y., Zaggia, S., Gilmore, G., Piotto, G., Carraro, G., Bedin, L.~R., \& de Angeli, F.\ 2006, \aap, 451, 515

\bibitem[Moni Bidin et al.(2009)]{2009RMxAC..35..109M} Moni Bidin, C., et al.\ 2009, Revista Mexicana de Astronomia y Astrofisica Conference Series, 35, 109

\bibitem[Moster et al.(2009)]{moster09} Moster, B.~P., Maccio',
A.~V., Somerville, R.~S., Johansson, P.~H.,
\& Naab, T.\ 2009, arXiv:0906.0764

\bibitem[Noguchi(1999)]{noguchi99} Noguchi, M.\ 1999, \apj, 514, 77

\bibitem[Ocvirk et al.(2008)]{ocvirk08} Ocvirk, P., Pichon, C., \& Teyssier, R.\ 2008, MNRAS,
390, 1326

\bibitem[Quinn et al.(1993)]{quinn93} Quinn, P.~J., Hernquist,
L., \& Fullagar, D.~P.\ 1993, \apj, 403, 74

\bibitem[Read et al.(2009)]{read09} Read, J.~I., Mayer, L.,
Brooks, A.~M., Governato, F., \& Lake, G.\ 2009, \mnras, 397, 44

\bibitem[Reddy et al.(2006)]{reddy06} Reddy, B.~E., Lambert,
D.~L., \& Allende Prieto, C.\ 2006, \mnras, 367, 1329

\bibitem[Robertson et al.(2006)]{robertson06} Robertson, B.,
Bullock, J.~S., Cox, T.~J., Di Matteo, T., Hernquist, L., Springel, V.,
\& Yoshida, N.\ 2006, \apj, 645, 986

\bibitem[Robin et al.(2003)]{robin03} Robin, A.~C., Reyl{\'e}, C.,
Derri{\`e}re, S., \& Picaud, S.\ 2003, \aap, 409, 523

\bibitem[Ro{\v s}kar et al.(2008)]{roskar} Ro{\v s}kar, R., Debattista, V.~P., Quinn, T.~R., Stinson, G.~S., \& Wadsley, J.\ 2008, \apjl, 684, L79 

\bibitem[Sch{\"o}nrich \& Binney(2009)]{schonrich-binney09}
Sch{\"o}nrich, R., \& Binney, J.\ 2009, \mnras, 396, 203

\bibitem[Seth et al.(2005)]{seth05} Seth, A.~C., Dalcanton, J.~J., \& de Jong, R.~S.\ 2005, \aj, 130, 1574

\bibitem[Shapiro et al.(2008)]{shapiro} Shapiro, K.~L., et al.\ 
2008, \apj, 682, 231 

\bibitem[Villalobos
\& Helmi(2008)]{VH08} Villalobos, {\'A}., \& Helmi, A.\ 2008, \mnras, 391, 1806

\bibitem[Walker et al.(1996)]{walker96} Walker, I.~R., Mihos,
J.~C., \& Hernquist, L.\ 1996, \apj, 460, 121

\bibitem[Yoachim
\& Dalcanton(2006)]{2006AJ....131..226Y} Yoachim, P., \& Dalcanton, J.~J.\ 2006, \aj, 131, 226 (YD06)

\bibitem[Zoccali et al.(2007)]{zoccali07} Zoccali, M., et al.\
2007, IAU Symposium, 241, 73

\end{thebibliography}
\end{document}